\documentclass[seceq]{ptptex}
\usepackage[dvips]{graphicx}
\usepackage{dcolumn}
\usepackage{bm}
\usepackage{psfrag}

\title{
 Detecting Generalized Synchronization of Chaotic Dynamical Systems
}

\subtitle{A Kernel-based Method and Choice of Its Parameter}    

\author{
Hiromichi \textsc{Suetani}$^{1}$, 
Yukito \textsc{Iba}$^{2}$, 
and Kazuyuki \textsc{Aihara}$^{3, 1}$
}

\inst{
$^{1}$Aihara Complexity Modelling Project, ERATO, JST, Tokyo 151-0064, Japan \\
$^{2}$The Institute of Statistical Mathematics, Tokyo 106-8569, Japan \\
$^{3}$Institute of Industrial Science, The University of Tokyo, Tokyo 153-8505, Japan
}

\abst{
An approach based on the {\it kernel methods} for capturing the nonlinear interdependence between two signals is introduced.
It is demonstrated that the proposed approach is useful for  characterizing generalized synchronization with a successful simple example. 
An attempt to choose an optimal kernel parameter based on {\it cross validation} is also discussed. 
}

\begin{document}

\maketitle

\section{Introduction}
Synchronization of chaotic systems has been explored extensively in recent years~\cite{Piko01}.
In addition to complete synchronization between two identical chaotic systems~\cite{Fuji83}, various notions of chaotic synchronization have evolved~\cite{Piko01}.
Among them,  the concept of {\it generalized synchronization} (GS), which  refers to a situation in which the states of two systems connected each other via a continuous mapping, has been introduced in order to study coherent behavior between two systems with different dynamics~\cite{Rulk95}.
Experimental detection of GS from data is a challenging problem.
Because the synchronization manifold of GS has a highly nonlinear structure, 
conventional statistical tools such as the correlation coefficient does not work.

Recently, the interest in the {\it kernel methods} has been stimulated in the machine learning community for analyzing data with nonlinearity in a unified manner~\cite{Shaw04}.
Since the great success of Support Vector Machine, a considerable effort has been devoted to derive kernelization of various multivariate analysis methods.
Therefore, it is meaningful to explore applicability of the kernel-based methods for analyzing nonlinear dynamics.  

In this paper, we particularly employ {\it Kernel Canonical Correlation Analysis (Kernel CCA)}~\cite{Akah01} for characterizing GS. 
We present an example for which Kernel CCA works successfully
and  also discuss how an optimal value of parameter of Kernel CCA can be chosen.  
 
\section{Kernel CCA}
Let us start with a formulation of Kernel CCA.
For a pair of variables $x\in {\mathbb R}^p$ and $y\in {\mathbb R}^q$, Kernel CCA seeks a pair of nonlinear scalar functions  $f: {\mathbb R}^p \to {\mathbb R}$ and $g: {\mathbb R}^q \to {\mathbb R}$ such that the correlation coefficient
\begin{eqnarray}
\label{rho_F}
\rho_{\cal F} = \frac{{\sf cov} (f(x), g(y))}{\sqrt{{\sf var}(f(x))}\sqrt{{\sf var}(g(y))}}
\end{eqnarray}
between transformed variables is maximized. 
When a data set $\{(x_n, y_n)\}_{n=1}^N$ is given, the maximal value of $\rho_{\cal F}$ is estimated from the following procedure. 

Suppose that the nonlinear functions $f$ and $g$ are well approximated by linear combinations of {\it kernels} on data points $(x_i, y_i)$ as
$f(x) = \sum_{i=1}^N \alpha_i k(x_i, x)$
and 
$g(y) = \sum_{i=1}^N \beta_i k(y_i, y)$.
For example, a Gaussian kernel $k(x, x') = \exp(-\Vert x - x'\Vert^2/2\sigma^2)$ is used as kernel~\cite{Shaw04}. 
By substituting the above expressions for $f$ and $g$ into Eq.~(\ref{rho_F}) and replacing covariance ${\sf cov}(\cdot,\cdot)$ and variance ${\sf var}(\cdot)$ with the empirical averages over the data set $\{(x_n, y_n)\}_{n=1}^N$, 
the maximization problem of Eq.~(\ref{rho_F}) leads to the following generalized eigenvalue problem~\cite{Akah01}: 
\begin{eqnarray}
\label{KCCA}
&&
 \left [
\begin{array}{cc}
0 & K_X K_Y \\
 K_Y K_X & 0
\end{array}
\right ]
\left[
\begin{array}{c}
\alpha \\
\beta
\end{array}
\right]
=
\rho
\left [
\begin{array}{cc}
K_X(K_X + \kappa I) & 0 \\
 0 & K_Y (K_Y + \kappa I)
\end{array}
\right ]
\left[
\begin{array}{c}
\alpha \\
\beta
\end{array}
\right],
\end{eqnarray}
where $K_X, K_Y$ are the {\it Gram matrices} $(K_X)_{i, j} = k(x_i, x_j)$ and $(K_Y)_{i, j} = k(y_i, y_j)$ determined from the given data set, and the term $\kappa I$ is introduced in order to avoid over-fitting.
The first eigenvalue of Eq.~(\ref{KCCA}) gives the maximal value $\rho_{\cal F}^{\max}$ of $\rho_{\cal F}$ in Eq.~(\ref{rho_F}).
$\rho_{\cal F}^{\max}$ is called the {\it canonical correlation coefficient}, and the variables $u=f(x), v=g(y)$ transformed by $f$ and $g$ are called the {\it canonical variates} of Kernel CCA.  

When the averages of $\{f(x_n)\}_{n=1}^N$ and $\{g(x_n)\}_{n=1}^N$  do not equal to zero, the Gram matrix $K$ should be replaced with the following one: 
$\tilde{K} = K - 
(1/N) ({\bf j} \> {}^t{\bf j}) K - 
(1/N) K ({\bf j} \>  {}^t{\bf j})  + 
(1/N^2) ({\bf j} \>  {}^t{\bf j}) K ({\bf j} \> {}^t{\bf j})$,
where ${\bf j} = {}^t (1,1,...,1)$~\cite{Shaw04}.  

\section{Results}
\begin{figure}[!t]
\psfrag{x1}{$x_1$}\psfrag{y}{$y$}
\psfrag{u}{$u$}\psfrag{v}{$v$}
\parbox{\halftext}{
\includegraphics[width=6.8cm]{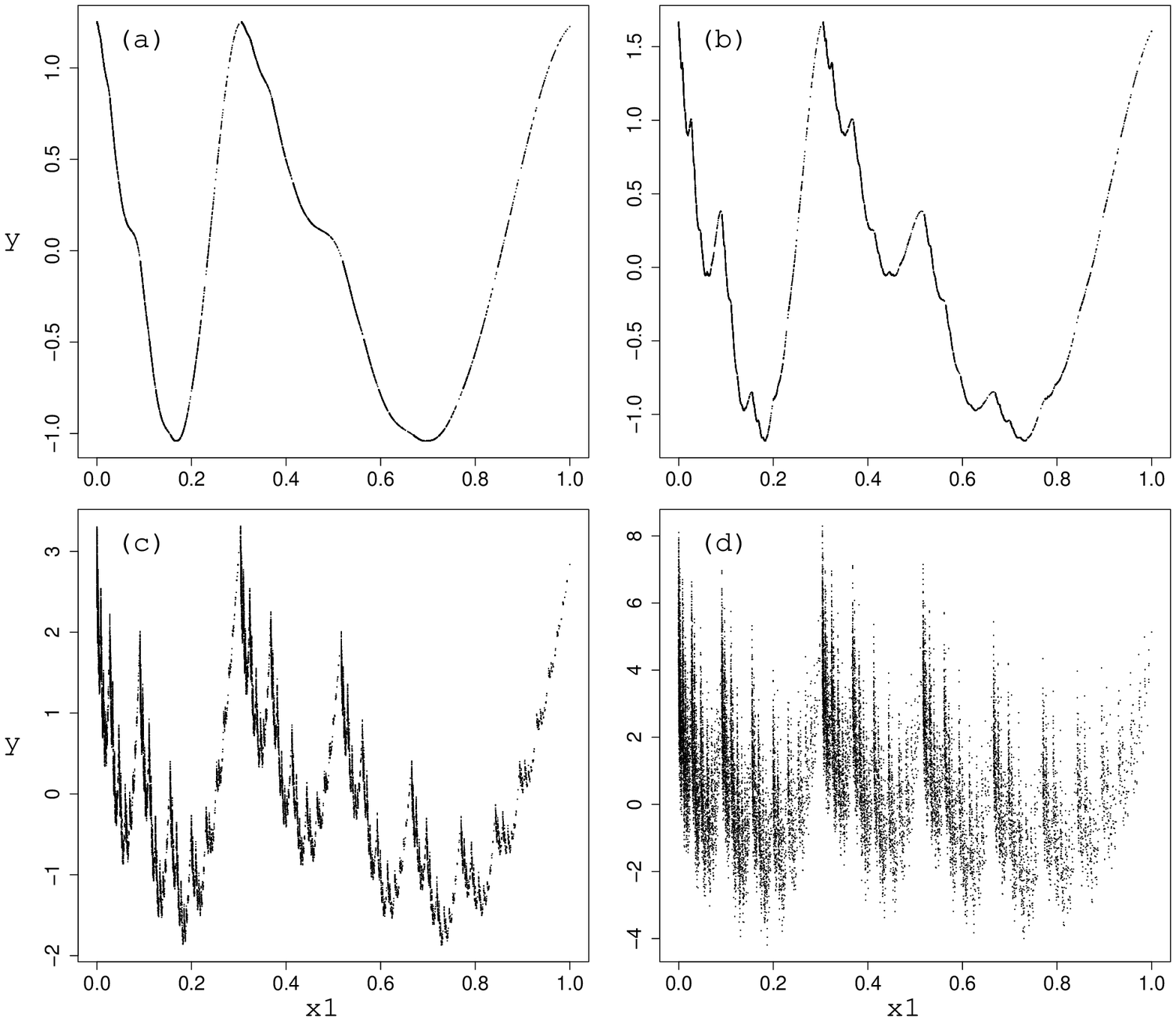}
\caption{Projections of the strange attractors onto the $(x_1, y)$ plane. $\gamma=0.2$ (a), $0.4$ (b), $0.7$ (c), $0.9$ (d).}
\label{fig.1}}
\ \ 
\parbox{\halftext}{
\includegraphics[width=6.4cm]{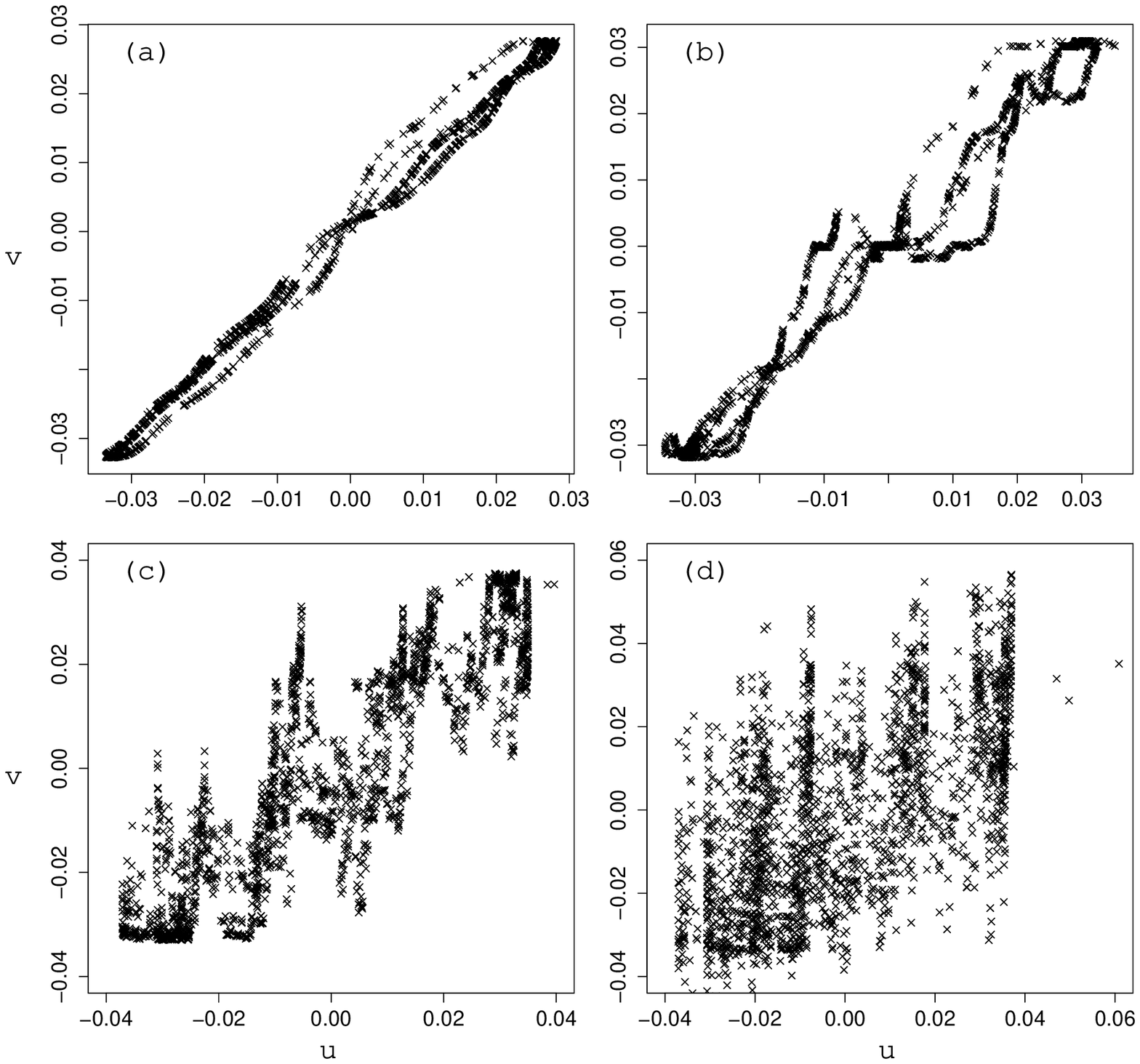}
\caption{Scatter plots of the first canonical variates of Kernel CCA. $\gamma=0.2$ (a), $0.4$ (b), $0.7$ (c), $0.9$ (d).}
\label{fig.2}}
\end{figure}

As an illustration, let us consider the following one-dimensional linear map driven by the two-dimensional baker's map: 
\begin{eqnarray}
\label{Baker}
 && [x_1(t+1), x_2 (t+1)]  =
\left \{
\begin{array}{lll}
[a x_1 (t), x_2 (t)/b]  \\ \quad \quad  {\rm if} \ x_2(t)<b, \\

[a + (1-a)x_1 (t), (x_2 (t) - b)/(1-b)] \\ \quad \quad  {\rm if} \  x_2(t) \geq b, 
\end{array}
\right.
\\
\label{Linear}
&& y (t+1)  =  \gamma y (t) + \cos(2\pi x_1 (t)).
\end{eqnarray}
The parameter of the baker's map Eq.~(\ref{Baker}) are taken as $a=0.3,  b=0.5$ and $\gamma$ in  Eq.~(\ref{Linear}) is varied as the control parameter.
For $|\gamma | < 1$, the response system Eq.~(\ref{Linear}) is asymptotically stable for all $(x_1, x_2)$ in the unit square $0\le x_1, x_2\le 1$, i.e., the system is in a state of GS.
Since the natural measure of the baker's map is uniform in the $x_2$ direction, the driver-response relation in the system of Eqs.~(\ref{Baker}) and (\ref{Linear}) is visualized in the $(x_1, y)$ plane as shown in Fig.~\ref{fig.1}. 
We observe the transition from smooth to very complicated curves with the increase of $\gamma$~\cite{Hunt97}. 

\begin{figure}[!t]
\psfrag{r}{$\gamma$}\psfrag{rho}{$\rho_{\cal F}^{\max}$}
\psfrag{DL}{$D_L$}\psfrag{sigma}{$\sigma$}
\parbox{\halftext}{
\includegraphics[width=6.2cm]{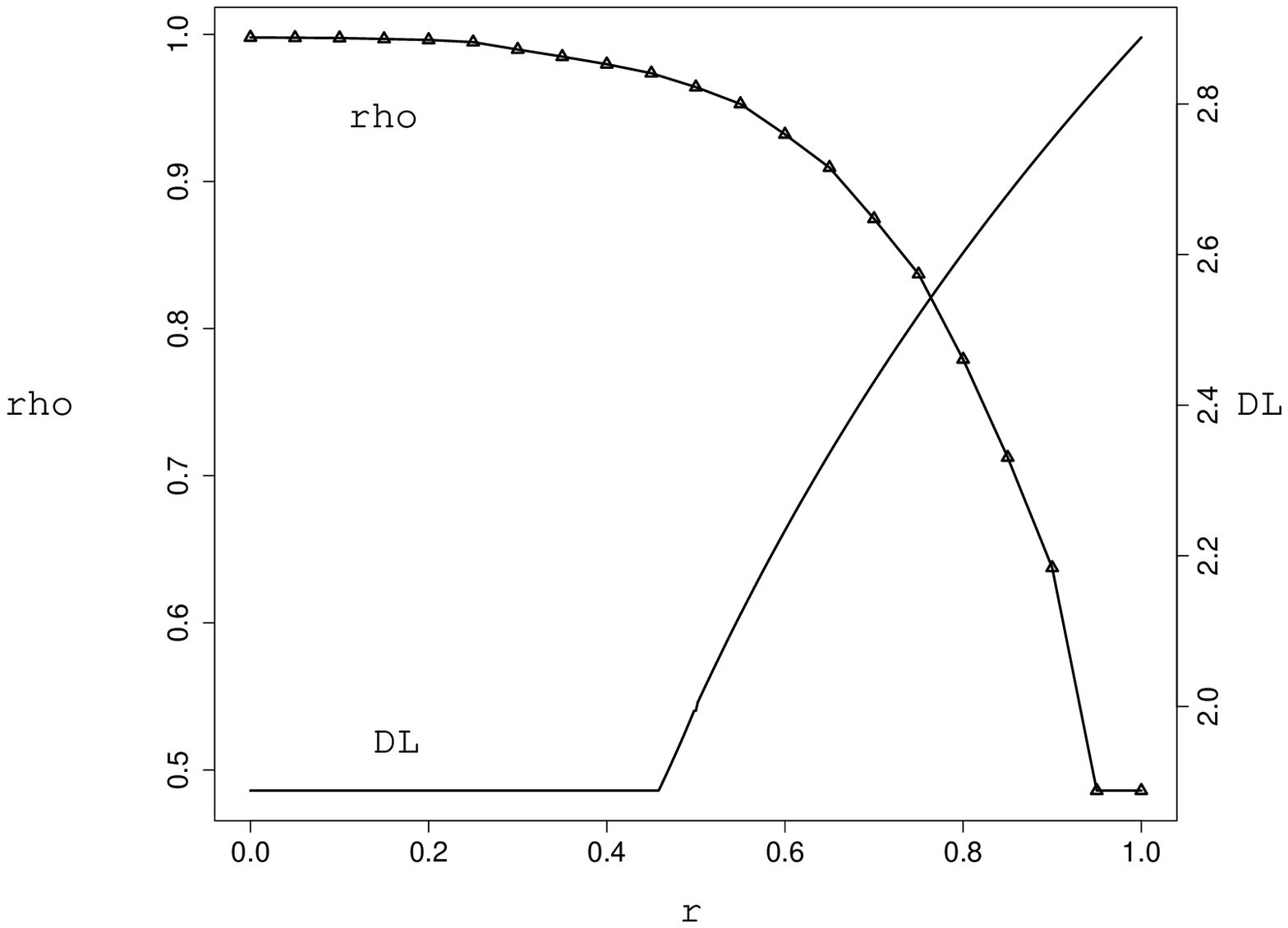}
\caption{The canonical correlation coefficient $\rho_{\cal F}^{\max}$ and the Lyapnov dimension $D_L$ as functions of $\gamma$. $N=2\times 10^3$, $\sigma=0.1$ and $\kappa=0.1$.}
\label{fig.3}}
\ \ 
\parbox{\halftext}{
\includegraphics[width=6.6cm]{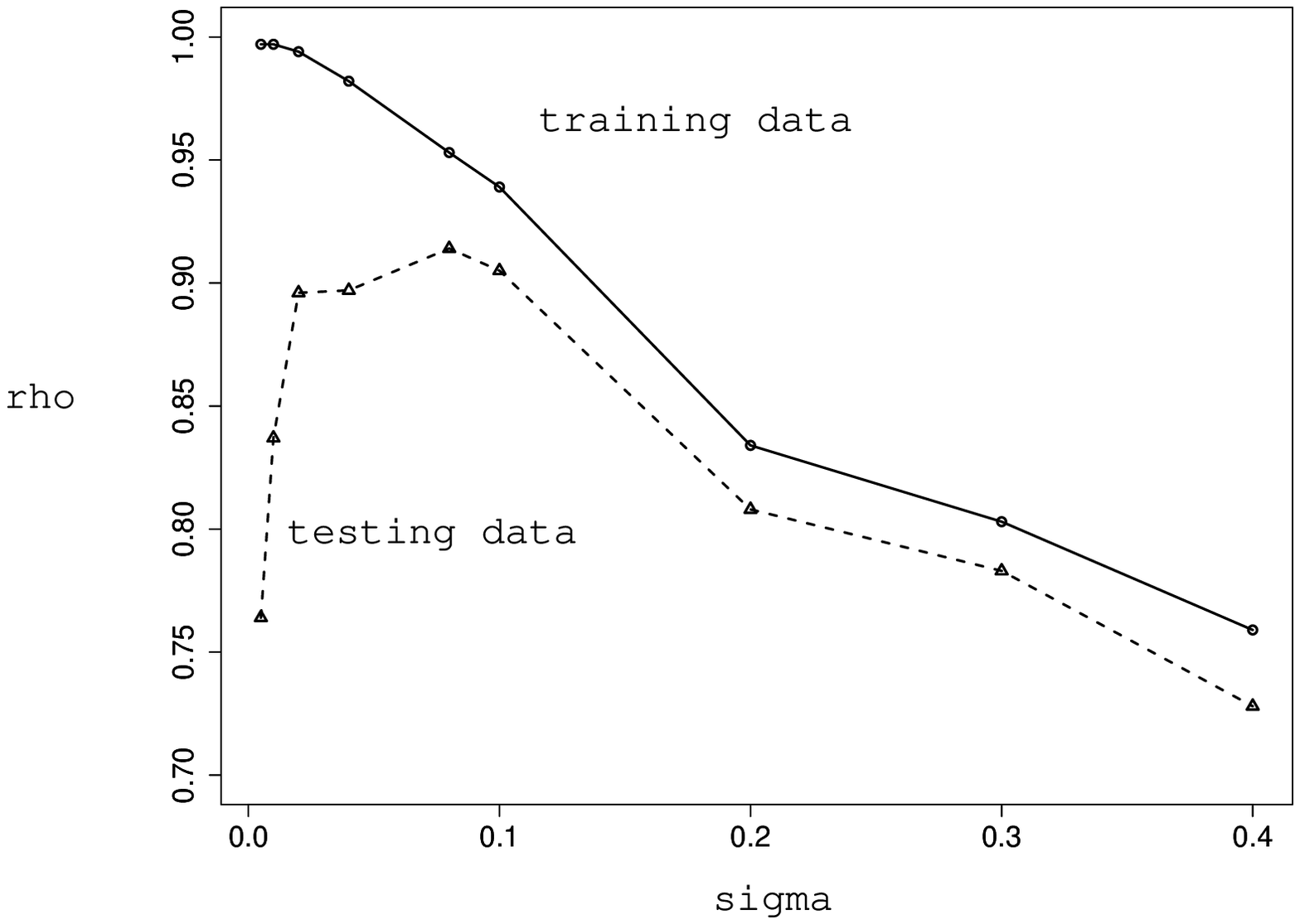}
\caption{$\rho_{\cal F}^{\max}$ as functions of $\sigma$ for $\gamma=0.6$. Orbits with length $N=2\times 10^2$ and $N'=10^4$ are used as training and testing data, respectively. $\kappa = 0.01$.}
\label{fig.4}}
\end{figure}

We apply Kernel CCA to the system of Eqs.~(\ref{Baker}) and (\ref{Linear}). 
Here, we employ a Gaussian kernel with $\sigma=0.1$ and $\kappa$ is set to $0.1$.
We prepare an orbit with length $N=2\times10^3$ as training data.  
Figure~\ref{fig.2} shows scatter plots of the canonical variates $(u_n, v_n)$ of Kernel CCA for four different values of $\gamma$ associated with Fig.~\ref{fig.1}.
In Fig.~\ref{fig.2}, GS is clearly identified as a cloud of points along the diagonal on the plane of the canonical variates.
By comparing graphs in Fig.~\ref{fig.2} with those in Fig.~\ref{fig.1},  the smaller value of the  correlation coefficient between canonical variates corresponds to the complexity of the structure of the synchronization manifold. 
Figure~\ref{fig.3} shows the canonical correlation coefficient $\rho_{\cal F}$ and the Lyapunov dimension $D_L$ of the system of Eqs.~(\ref{Baker}) and (\ref{Linear})  as functions of the control parameter $\gamma$.
There is a monotonic relationship between $\rho_{\cal F}^{\max}$ and $D_L$, which allows us to conclude that  $\rho_{\cal F}^{\max}$ is a good index for 
characterization of GS. 

Next, we mention how the parameters of Kernel CCA can be chosen from data. 
In order to represent complicated nonlinear structures of the synchronization manifold   via  linear combinations of Gaussian kernels, the value of $\sigma$ should be chosen adequately.

A naive way of choosing $\sigma$ is to maximize $\rho_{\cal F}^{\max}$ estimated by the proposed method. 
The estimator of $\rho_{\cal F}^{\max}$ as a function of $\sigma$ is shown with the solid line in Fig.~\ref{fig.4}. 
The index $\rho_{\cal F}^{\max}$ increases monotonically with the decrease of $\sigma$, and $\rho_{\cal F}^{\max} \sim 1$ is attained in the limit of $\sigma\to 0$.
Thus the maximization of
$\rho_{\cal F}^{\max}$ evaluated from the training data leads to the choice of the smallest value of $\sigma$, which results in over-fitting to the training data and   spurious detection of GS.
An adequate value of $\sigma$ is not determined in this way.

A way to overcome this difficulty is to prepare another set of data (``testing data") separately from the training data, and evaluate $\rho_{\cal F}^{\max}$ from the empirical average over the testing data, while $f$ and $g$ are estimated from the training data.
This strategy for testing the performance of the estimated model  with new data is called {\it cross validation (CV)}~\cite{Ston74}.
The dotted line in Fig.~\ref{fig.4} shows the result of CV. 
The value of  $\rho_{\cal F}^{\max}$ shown with the dotted line takes its maximum at a value of  $\sigma\neq 0$.  
We expect that this value of $\sigma$ gives an optimal description of the system behind data.
\section{Conclusions}
In summary, we have proposed a new approach for analyzing GS based on Kernel CCA with a successful application to a simple example.
It is interesting to apply other kernel-based methods for analyzing various complex phenomena arising from nonlinear dynamical systems.  

\section*{Acknowledgements}
We thank S.~Akaho  and K.~Fukumizu for stimulating discussions on the kernel methods. 

%

\end{document}